\documentclass[a4paper,10pt, english]{article}    

\usepackage[a4paper,left=1cm,right=1cm,top=1cm,bottom=1.5cm,bindingoffset=5mm]{geometry}
\usepackage{multicol}
\usepackage[utf8]{inputenc}               
\usepackage[T1]{fontenc}                  
\usepackage[autostyle=true]{csquotes}     
\usepackage{graphicx}                     
\usepackage{wrapfig}                      
\usepackage{float}					      
\usepackage{amsmath}                      
\usepackage{amssymb}                      
\usepackage{commath}                      
\usepackage{xfrac}                        
\usepackage{IEEEtrantools}                
\usepackage{siunitx}                      
\usepackage[varg]{txfonts}                
\usepackage{hyperref}                     
\usepackage[all]{hypcap}                  
\usepackage{enumitem}                     
\setenumerate{leftmargin=*}               
\graphicspath{{figs/}}                    
\usepackage{tikz}
\usepackage{lipsum}
\usepackage[backend=biber,style=numeric,sortcites,sorting=none,natbib,hyperref]{biblatex}
\usepackage[toc, page]{appendix}
\usepackage[figuresright]{rotating}
\usepackage{makecell}
\usepackage{indentfirst}
\usepackage{authblk}
\usepackage{caption}
\usepackage{dblfnote}
\makeatletter
\title{ The possible emergence of an attractive inverse-square law from the wave-nature of particles}
\author[1,*]{Dong Zhang}
\author[1]{Pavel Kroupa}
\author[1]{Jan Pflamm-Altenburg}
\author[2]{Manfred Schmid}
\affil[1]{Helmholtz-Institut für Strahlen und Kernphysik, Universität Bonn, Nussallee 14-16, Bonn, 53115, Germany.}
\affil[2]{Eboracum GmbH, Im Vogelsang 9, Wachtberg, 53343, Germany.}
\affil[*]{Corresponding author: Dong Zhang, \href{mailto:s6dozhan@uni-bonn.de}{s6dozhan@uni-bonn.de}}
\date{\today}
\begin{document}
	\maketitle	
	
	\begin{abstract}
		A model of a particle in finite space is developed and the properties that the particle may possess under this model are studied.
		The possibility that particles attract each other due to their own wave nature is discussed. The assumption that the particles are spatially confined oscillations (SCO) in the medium is used. The relation between the SCO and the refractive index of the medium in the idealized universe is derived. Due to the plane wave constituents of SCOs, the presence of a refractive index field with a non-zero gradient causes the SCO to accelerate. The SCO locally changes the refractive index such that another SCO is accelerated towards it, and vice versa. It is concluded that the particles can attract each other due to their wave nature and a inverse-square-type acceleration emerges. 
		The constant parameter in the inverse-square-type acceleration is used to compare with the gravitational constant $G_{N}$, and the possibility of non inverse-square-type behavior is preliminary discussed.
		
	\end{abstract}

	\begin{multicols}{2}
		
		\section{Introduction}
		\label{intro}
		It is generally accepted in modern physics that there are four known fundamental interactions: gravitation, electromagnetism, the weak and the strong interaction. In the 1960s, Glashow \cite{Glashow1961PartialSO}, Salam \cite{Salam:1968rm} and Weinberg \cite{weinberg1967model}, gave a theory of electroweak unification that has been experimentally verified (Hasert et al. \cite{article,GargamelleNeutrino:1974khg}, Arnison et al. \cite{UA1:1983mne,UA1:1983crd}, etc).  In modern times, the standard model of particle physics is a theory based on the description of three interactions except for gravitation. In 2012, experiments at the large hadron collider (LHC) confirmed the existence of the Higgs boson (Aad et al.\cite{ATLAS:2012yve}), indicating that all particles described by the Standard Model were confirmed to exist. But the standard model still has difficulties in explaining some phenomena, for example, it does not explain the existence of dark matter and dark energy at all, and it neither describes gravitation nor mass.
		
		There exist some theories that try to unify the Standard Model and general relativity, such as quantum gravity. These approaches have not yet lead to significant progress. Rather than seeking a model that incorporates dark matter particles that have not been experimentally verified (Kroupa \cite{kroupa_2012,Kroupa:2014ria}), another approach is to improve our gravitational theory. 
		A modification of gravitational theory has been proposed by Milgrom in 1983 \cite{Milgrom1983AMO} to explain the phenomena not compatible with Newtonian mechanics in astronomy (Milgromian Dynamics, MOND).

		In 2020, Stadtler et al. \cite{stadtler2021dynamics} proposed the spatially confined oscillation (SCO) model of particles in an idealized  universe based on the wave nature of matter and tried to use it to study the gravitational problem. This concept is an extension of an earlier article (Schmid and Kroupa \cite{schmid_kroupa_2014}), specifically it extends the concept of a spheron, i.e., standing spherical wave, to a generalised spatially confined oscillation. In the SCO model of particles, the particle is considered as a superposition of standing waves, where the waves are considered as propagating perturbations of the medium filling in the idealized universe. The medium in this idealized universe can be considered to be an ideal classical physical medium. The propagation speed of the wave in the medium is $c_{s}$. In the theory, an SCO is accelerated by being affected by a refractive index field, $n_{ref}$, with a non-zero gradient. 
		
		For gravitation, Newton's law of universal gravitation has a good accuracy in weak gravitational fields, at low speeds and at long distances, especially in astronomical measurements, and it is an inverse-square law. 
		In this contribution, we will discuss the possibility of spontaneous mutual attraction between SCOs from the SCO mechanism and try to develop a inverse-square-type ($1/r^{2}$-type, $r$ is the distance) attraction model. Then we will study the possible properties of the SCO under this model.
		
		In Section 2 the SCO model and the equations of motion of the SCO are presented in detail. The relation between the refractive index and the properties of the medium itself, such as pressure and density in fluid dynamics, are introduced. In Section 3 we try to find an expression for the refractive index field of the medium through the equation of motion of an SCO so that the SCO has a inverse-square-type acceleration. In Section 4 we construct SCO models and combine them with fluid dynamics to study refractive index changes spawned by an SCO and we search for a inverse-square-type acceleration. A model is found for the superposition of multiple-spherical standing waves and a inverse-square-type acceleration is obtained. The possibility of non inverse-square-type behavior is preliminarily discussed. Section 5 provides a summary and outlook for the further research.

	\section{Theoretical basis}
	\subsection{Model of SCO and the equation of motion}
	\label{SCOModel}
		Stadtler et al. \cite{stadtler2021dynamics} gave the full definition of an SCO. Here is an introduction to the SCO model and the equation of motion.
		
		The SCO model for a particle means that the particle is assumed to be a spatially confined oscillation. It is a primitive particle as it only carries energy and is not associated with other properties like spin, charge, etc. Waves superimposed into this kind of periodic oscillation can be interpreted as the propagation of a spatially confined perturbation in a homogeneous, isotropic medium with the propagation speed $c_{s}$, i.e., a mechanical wave, such as a fluctuation in pressure or density. In a more general case, this oscillating field of an SCO, $\rho(\textbf{x},t)$, can have a more complex relation with the density field $D(\textbf{x},t)$ and the pressure field $P(\textbf{x},t)$ of the medium. In this contribution we consider one intuitive case: The oscillatory field of the SCO is proportional to the density field or pressure field of the medium, i.e., $D(r,t) = \alpha_{D} \rho(r,t)$ or $P(r,t) = \alpha_{P} \rho(r,t)$ with $\alpha_{D}$ and $\alpha_{P}$ being constants. Here $\textbf{x}$ is the 3-dimensional spatial coordinate and $t$ is the time coordinate. The SCO can transfer energy through the medium.
		
		If the oscillation satisfies the wave equation (to make the computations tractable we limit the discussion here to the classical wave equation which is valid only for a constant $c_{s}$. We relax this by considering the relative change in $c_{s}$, $\delta c_{s} / c_{s} \ll 1$, in order to assess if two SCOs might attract each other as a consequence of the changed refractive index without ensuring full consistency of the SCO solution of the wave equation. A self-consistent analysis will need to take into account solitons (e.g., Rajaraman \cite{rajaraman1982solitons}).)
		  \begin{equation}\label{waveeq}
          \frac{\partial^{2} \rho(\textbf{x},t)}{\partial t^{2}} = c^{2}_{s} \nabla ^{2} \rho(\textbf{x},t),
          \end{equation}
		the Fourier transformation can be used to obtain the plane-wave superposition representation of the oscillation
		  \begin{equation}\label{fourtr}
          \rho(\textbf{x},t) = Re\left[\int\mathrm{d}^{3}k  A(\textbf{k}) \mathrm{e}^{i\textbf{k}\cdot\textbf{x} - i\omega t } \right].
          \end{equation}
        
        Here $\omega $ is the time frequency of the oscillation and the non-dispersive relation $\omega = c_{s}|\textbf{k}| = c_{s}k$ holds. $\textbf{k}$ is the wave vector of a plane wave with $|\textbf{k}| = 2\pi/\lambda = k$. It expresses the wavelength $\lambda$ and direction of motion of a plane wave. $A(\textbf{k})$ is a function of $\textbf{k}$.
		
		In the description of \cite{stadtler2021dynamics}, a single plane wave remains a solution of the wave equation, Eq.(\ref{waveeq}), after the active Lorentz transformation. Thus after the active Lorentz transformation, an oscillation as a whole still can be expressed as a superposition of plane waves. This means that the SCO remains the solution of the wave equation and could explain why special relativity arises due to the wave nature of matter particles.
		
		The active Lorentz transformation has physical significance. An inhomogeneous medium has a varying propagation speed field $c_{s}(\textbf{x},t)$, or refractive index field $n_{ref}(\textbf{x},t) = c_{s,0}/c_{s}(\textbf{x},t)$, where $c_{s,0}$ is considered to be the propagation speed possessed by the homogeneous ambient medium. When a plane wave propagates in an inhomogeneous medium, the effect on the plane wave due to changes in the refractive index can be described by the active Lorentz transformation.
		
		As described by \cite{stadtler2021dynamics}, the gradient of propagation speed experienced by the SCO can be considered constant when the scale of the SCO is much smaller than the spatial scale over which the change of the propagation speed is apparent. Without loss of generality, set the center of the SCO at the origin of the spatial coordinates. Then the propagation speed around the SCO is expanded as:
		  \begin{equation}\label{spdgra}
		  \begin{aligned}
		  c_{s}(\textbf{x}) &= c_{s}(0) + \nabla c_{s}(\textbf{x})|_{\textbf{x} = \textbf{0}} \cdot \textbf{x} + \mathcal{O}(|\textbf{x}|^{2})  \\
		  & = c_{s,0} + \textbf{c}'_{s} \cdot \textbf{x} + \mathcal{O}(|\textbf{x}|^{2}),
		  \end{aligned}
          \end{equation}
        where $\textbf{c}'_{s}$ is in the direction of increase in propagation speed.
        
		The Lorentz transformation along the $\textbf{n}$ direction and in time coordinate can be written as:
		  \begin{equation}\label{spacelt}
		  \begin{aligned}
          & \textbf{x}'(\textbf{x},t) = \gamma(\textbf{x} \cdot \textbf{n} - \beta c_{s} t)\textbf{n} + \textbf{x} - \textbf{n}(\textbf{x} \cdot \textbf{n}),\\
          & c_{s}t'(\textbf{x},t) = \gamma(c_{s}t - \beta \textbf{x}\cdot\textbf{n}).
          \end{aligned}
          \end{equation}
        
        Here $\gamma$ is the Lorentz factor, $\beta = v/c_{s}$. And $\gamma = 1/\sqrt{1 - \beta^{2}}$. This coordinate transformation can also be written in matrix form for convenience:
		  \begin{equation}\label{spaceltmtr}
          \begin{bmatrix}
          \gamma & -\beta \gamma \textbf{n}^{T} \\ 
          -\beta \gamma \textbf{n} & \textbf{I} + (\gamma - 1)\textbf{n}\textbf{n}^{T} 
          \end{bmatrix} 
          \begin{pmatrix}
          c_{s}t  \\ 
          \textbf{x}
          \end{pmatrix} = \textbf{R}(\gamma,\textbf{n}) 
          \begin{pmatrix}
          c_{s}t  \\ 
          \textbf{x}
          \end{pmatrix}=
          \begin{pmatrix}
          c_{s}t'  \\ 
          \textbf{x}'
          \end{pmatrix},
          \end{equation}
        where \textbf{I} is the $3 \times 3$ identity matrix, $\textbf{n}^{T} $ is the transpose of the vector \textbf{n}, $\textbf{R}(\gamma,\textbf{n})$ is the Lorentz transformation matrix and represents the Lorentz transformation along the $\textbf{n}$-direction with Lorentz factor $\gamma$ which contains information about the velocity of the transformation.
        
        The plane wave is affected by the propagation speed field $c_{s}(\textbf{x},t)$ with non-zero gradient $\nabla c_{s}(\textbf{x})$ as it passes through an inhomogeneous medium. According to the derivation of \cite{stadtler2021dynamics}, the differential equation of the time rate of change of the wave vector is:
          \begin{equation}\label{snell}
          \dot{\textbf{k}}(t) = - k(t)\textbf{c}'_{s}.
          \end{equation}
        This indicates that, as time evolves, the wave vector $\textbf{k}$ will tend to be directed in the direction of decreasing propagation speed, i.e., the direction of increasing refractive index of the medium. A plane wave remains a solution to the wave equation, Eq.(\ref{waveeq}), after the Lorentz transformation. According to Eq.(\ref{fourtr}), this means that a plane wave (i.e., a constituent of the SCO) undergoes the following transformation:
		  \begin{equation}\label{showlt}
          e^{i(\textbf{k}\cdot \textbf{x} - kc_{s,0}t)} \rightarrow e^{i(\textbf{k}\cdot \textbf{x}' - kc_{s,0}t'  )}.
          \end{equation}
        Since with the description in \cite{stadtler2021dynamics}, the amplitudes and phase offsets of all plane wave constituents remain the same, so that the function $A(\textbf{k})$ remains the same in the integration over the frequency domain and only the exponential term, i.e., wave part shown in Eq.(\ref{showlt}), is transformed.
        
		The active Lorentz transformation will keep the form of the plane wave unchanged, so the wave vector of the plane wave constituent satisfies the following relation:
		  \begin{equation}\label{wavevecre}
          \textbf{k}'\cdot \textbf{x} - k'c_{s,0}t = \textbf{k}\cdot \textbf{x}' - kc_{s,0}t'.
          \end{equation}
          
        Eq.(\ref{spacelt}) is then used to replace $\textbf{x}'$ and $c_{s,0}t'$, so that the representation of the wave vector, $\textbf{k}'$, of the plane wave after the Lorentz transformation by $\textbf{k}$ is obtained. $\textbf{k}'$ is represented as
          \begin{equation}\label{kspacelt}
		  \begin{aligned}
          &\textbf{k}' = \gamma'(t)[\textbf{k} \cdot \textbf{e}' - \beta'(t) k]\textbf{e}' + \textbf{k} - \textbf{e}'(\textbf{k} \cdot \textbf{e}'),&\\
          &k' = \gamma'(t)[k - \beta'(t) \textbf{k}\cdot\textbf{e}'].
          \end{aligned}
          \end{equation}
        Here $k'$ is the wave number after the transformation and  $k' = |\textbf{k}'|$. The forms $\gamma'(t) = \cosh{(c'_{s}t)}$ and $\beta'(t) = \tanh{(c'_{s}t)}$ with $c'_{s} = |\textbf{c}'_{s}|$ are used. $\textbf{e}'$ is the unit direction vector of the gradient of propagation speed, i.e., $\textbf{c}'_{s} = c'_{s}\textbf{e}'$. A point to note is that the transformation of the wave vector, according to Eq.(\ref{snell}), indicates that the active Lorentz transformation is in the direction of the decrease in the propagation speed, i.e., the $-\textbf{e}'$ direction. Since the speed of propagation of a plane wave under a non-zero gradient field, i.e., $c'_{s} \neq 0$, will be related to the time of its motion in the medium, the Lorentz transformation is therefore also related to time.
        
        The SCO can be represented as a superposition of plane waves. When it is in the inhomogeneous medium as a whole and undergoes the Lorentz transformation, we have:
          \begin{equation}\label{superlt}
		  \begin{aligned}
          &\rho(\textbf{x},t) = Re\left[\int\mathrm{d}^{3}k  A(\textbf{k}) \mathrm{e}^{i\textbf{k}\cdot\textbf{x} - ikc_{s,0}t } \right] \rightarrow \\
          &Re\left[\int\mathrm{d}^{3}k  A(\textbf{k}) \mathrm{e}^{i\textbf{k}'\cdot\textbf{x} - ik'c_{s}[\lambda(t)]t } \right] = \rho'(\textbf{x},t).
          \end{aligned}
          \end{equation}
        That means a new representation of the same SCO can be obtained. In the description of \cite{stadtler2021dynamics}, $\lambda(t)$ is the trajectory of the centre of the SCO. From here on we use the movement of the centre of SCO to represent the movement of the SCO as a whole.
        
        Now according to Eq.(\ref{kspacelt}) we can replace the $\textbf{k}'$ and $k'$ of $\rho'(\textbf{x},t)$ in Eq.(\ref{superlt}) by $\textbf{k}$ and $k$  so that we get the form of the time and space coordinates after going through an inhomogeneous medium and the active Lorentz transformation. That means:
          \begin{equation}\label{spacexlt}
		  \begin{aligned}
          &\textbf{x}'(\textbf{x},t) = \gamma'(t)\{ \textbf{x} \cdot \textbf{e}' + \beta'(t) c_{s}[\lambda(t)] t\}\textbf{e}' + \textbf{x} - \textbf{e}'(\textbf{x} \cdot \textbf{e}'),&\\
          &c_{s,0}t'(\textbf{x},t) = \gamma'(t)\{c_{s}[\lambda(t)]t + \beta'(t) \textbf{x}\cdot\textbf{e}'\}.
          \end{aligned}
          \end{equation}
        Comparing with Eq.(\ref{spacelt}), it can be seen that the direction of the spatial coordinate transformation under the influence of inhomogeneous medium is $-\textbf{e}'$. In the transformation of the time coordinate, a new factor $c_{s}[\lambda(t)]/c_{s,0}$ is added to the time term of $t'$: $\gamma'(t)c_{s}[\lambda(t)]t/c_{s,0}$,   because the propagation speed is no longer constant in the inhomogeneous medium. The above transformation in Eq.(\ref{spacexlt}) is written in matrix form as
          \begin{equation}\label{spacexltmtr}
          \begin{aligned}
          &\begin{Bmatrix}
          \gamma'(t) & \beta'(t) \gamma'(t) \textbf{e}'^{T} \\ 
          \beta'(t) \gamma'(t) \textbf{e}' & \textbf{I} + [\gamma'(t) - 1]\textbf{e}'\textbf{e}'^{T} 
          \end{Bmatrix} 
          \begin{Bmatrix}
          c_{s}[\lambda(t)]t  \\ 
          \textbf{x}
          \end{Bmatrix} &\\ 
          &= \textbf{R}[\gamma'(t),-\textbf{e}'] 
          \begin{Bmatrix}
          c_{s}[\lambda(t)]t  \\ 
          \textbf{x}
          \end{Bmatrix}=
          \begin{pmatrix}
          c_{s,0}t'  \\ 
          \textbf{x}'
          \end{pmatrix}.
          \end{aligned}
          \end{equation}
        
        After the above description, we can now introduce the equation of motion of the SCO derived in \cite{stadtler2021dynamics}. The initial SCO can be represented as a Lorentz transformation of a stationary SCO: 
          \begin{equation}\label{initrest}
		  \begin{aligned}
          \rho_{init}(\textbf{x}',t') = \rho_{stat}[\textbf{x}''(\textbf{x}',t'),t''(\textbf{x}',t')].
          \end{aligned}
          \end{equation}
        Here $\rho_{init}$ is the oscillation field of the initial SCO, $\rho_{stat}$ is the oscillation field of the stationary SCO. The initial SCO can have a non-zero initial velocity, so a representation of the initial SCO can be obtained by performing a Lorentz transformation on a stationary SCO.
        
        The SCO, after being affected by the inhomogeneous medium, can be expressed as a Lorentz transformation of the initial SCO. 
          \begin{equation}\label{initfinal}
		  \begin{aligned}
          \rho_{final}(\textbf{x},t) = \rho_{init}[\textbf{x}'(\textbf{x},t),t'(\textbf{x},t)].
          \end{aligned}
          \end{equation}
        Here $\rho_{final}$ is the oscillation field of the SCO after being affected by the inhomogeneous medium, i.e., the final state of the SCO. Then according to Eq.(\ref{spaceltmtr}) and Eq.(\ref{spacexltmtr}) the transformation from the final state SCO to the stationary SCO can be represented by the matrix (vice versa):
          \begin{equation}\label{initfinalmtx}
          \begin{aligned}
          \textbf{R}[\gamma,\textbf{n}] \textbf{R}[\gamma'(t),-\textbf{e}'] 
          \begin{Bmatrix}
          c_{s}[\lambda(t)]t  \\ 
          \textbf{x}
          \end{Bmatrix}=
          \textbf{R}[\gamma,\textbf{n}]
          \begin{pmatrix}
          c_{s,0}t'  \\ 
          \textbf{x}'
          \end{pmatrix} = \begin{pmatrix}
          c_{s,0}t''  \\ 
          \textbf{x}''
          \end{pmatrix}.
          \end{aligned}
          \end{equation}
        $\textbf{R}[\gamma'(t),-\textbf{e}'] $ expresses the transformation matrix between the final state and initial SCO, along the direction of the decrease in the propagation speed, i.e, $-\textbf{e}'$. And the information on speed is expressed by $\gamma'(t)$. $\textbf{R}[\gamma,\textbf{n}]$ expresses the transformation matrix between the stationary and initial SCO. The initial SCO possesses (compared to the stationary SCO) motion along the $\textbf{n}$ direction. And the information on speed is expressed by $\gamma$.
        The transformation from the initial SCO to the final state SCO can also be expressed as a combination of one Lorentz transformation and one rotation, as shown by the concept of Thomas-Wigner rotation in \cite{Thomas:1926dy,Wigner:1939cj}:
          \begin{equation}\label{towi}
          \begin{aligned}
          \begin{pmatrix}
          c_{s,0}t''  \\ 
          \textbf{x}''
          \end{pmatrix} = 
          \begin{bmatrix}
          1 & 0  \\ 
          0 & \textbf{O}(\theta)
          \end{bmatrix} 
          \textbf{R}(\gamma_{new}, \textbf{n}_{new})
          \begin{Bmatrix}
          c_{s}[\lambda(t)]t  \\ 
          \textbf{x}
          \end{Bmatrix}.
          \end{aligned}
          \end{equation}
        Here $\textbf{O}(\theta)$ is the spatial rotation matrix with the rotation angle $\theta$. An introduction of Thomas-Wigner rotation is also given in \cite{stadtler2021dynamics}. After Thomas-Wigner rotation, the new Lorentz transformation is along the $\textbf{n}_{new}$-direction with the new Lorentz factor $\gamma_{new}$. After comparing the matrix $\textbf{R}[\gamma,\textbf{n}] \textbf{R}[\gamma'(t),-\textbf{e}'] $ in Eq.(\ref{initfinalmtx}) with the matrix $\textbf{R}(\gamma_{new}, \textbf{n}_{new})$ in Eq.(\ref{towi}), $\textbf{n}_{new}$ and $\gamma_{new}$ have the following expressions:
          \begin{equation}\label{towiga}
          \begin{aligned}
          &\gamma_{new}(t) = \gamma\gamma'(t)[1 - \beta\beta'(t)(\textbf{n}\cdot \textbf{e}')],&\\
          &\textbf{n}_{new}(t) = \frac{\beta\gamma\textbf{n} + [\beta\gamma(\gamma'(t)-1)(\textbf{n}\cdot \textbf{e}') - \gamma\gamma'(t)\beta'(t)  ] \textbf{e}'}{\beta_{new}(t)\gamma_{new}(t)}.
          \end{aligned}
          \end{equation}
        Here $\gamma_{new}$ contains information on the magnitude of the transformation velocity and $\textbf{n}_{new}$ indicates the direction of the transformation velocity. Since the Lorentz transformation acts on the whole reference system, $\gamma_{new}$ and $\textbf{n}_{new}$ describe the velocity of the SCO.
        
        Since $\textbf{n}_{new}$ is a unit direction vector, we can define the relative velocity vector as $\boldsymbol{\beta}_{new}(t) = \beta_{new}(t)\textbf{n}_{new}(t)$. Then we do the time differentiation for $\boldsymbol{\beta}_{new}(t)$ and obtain:
          \begin{equation}\label{betatime}
          \begin{aligned}
          \dot{\boldsymbol{\beta}}_{new}(t) = -\textbf{c}'_{s} + \boldsymbol{\beta}_{new}(t)[\textbf{c}'_{s} \cdot \boldsymbol{\beta}_{new}(t)].
          \end{aligned}
          \end{equation}
        This result can be extended to the propagation speed field with varying gradient, i.e., $\textbf{c}'_{s} \rightarrow \textbf{c}'_{s}[\lambda(t)]$. And it is also possible to translate the above result in Eq.(\ref{betatime}) into an expression for acceleration $\textbf{a}(t)$:
          \begin{equation}\label{accetime}
          \begin{aligned}
          \textbf{a}(t) = -c_{s}[\lambda(t)] \textbf{c}'_{s}[\lambda(t)] + 2c_{s}[\lambda(t)]\boldsymbol{\beta}_{new}(t)[\textbf{c}'_{s}[\lambda(t)] \cdot \boldsymbol{\beta}_{new}(t)].
          \end{aligned}
          \end{equation}
        Observing Eq.(\ref{accetime}), we find that the SCO has an acceleration that is related to its own velocity, and to the propagation speed field.  If the gradient of the propagation speed field is zero, then the SCO will have a constant velocity.  In other words, an SCO will acquire an acceleration in a changing propagation speed field, i.e., in a refractive index field.
          
        To derive the result of Eq.(\ref{accetime}) the following relation was used:  
          \begin{equation}\label{relati}
          \begin{aligned}
          \dot{\boldsymbol{\beta}}_{new}(t) = \frac{\mathrm{d}\left\{\frac{\textbf{v}(t)}{c_{s}[\lambda(t)]}\right\}}{\mathrm{dt}} = \frac{\textbf{a}(t)}{c_{s}[\lambda(t)]} - \frac{c_{s}[\lambda(t)]\boldsymbol{\beta}_{new}(t)[\textbf{c}'_{s}(t) \cdot \boldsymbol{\beta}_{new}(t)]}{c_{s}[\lambda(t)]}.
          \end{aligned}
          \end{equation}
          
        Up to now, we have obtained the equation of motion of the SCO as a whole in a propagation speed field or refractive index field with a non-zero gradient. Stadtler et al. \cite{stadtler2021dynamics} also point out that the SCO follows a geodesic equation such that the refractive index field can alternatively be interpreted as a space-time distribution in general relativity.
        
    \subsection{Derivation of refractive index in fluid dynamics}
    \label{sec2.2}    
        The idealized universe in the SCO model is filled with ideal classical physical matter, and we can assume that this matter possesses the properties of a compressible fluid. The wave is therefore defined as the propagation of perturbation in a fluid. Since the non-dispersive relation $\omega = |\textbf{k}|c_{s}$ is used, the phase velocity of a plane wave is equal to the group velocity. And the group velocity is the speed of propagation of the energy and information. So we need to study the propagation speed of a plane wave in the fluid. The expression of the phase velocity in an inhomogeneous medium has been described by Simaciu et al. \cite{Simaciu:2018ywl}. Here is an introduction of this expression.
        
        The wave changes the density and pressure of the fluid medium locally. And the motion of waves in an ideal fluid can be approximated as an adiabatic process. According to Landau and Lifshitz in \cite{landau2013fluid}, changes in the pressure and density of the medium affect the phase velocity of the plane wave. This means that the oscillation of pressure and density itself affects the phase velocity of nearby plane waves and causes a non-zero gradient in the local propagation speed field. The SCO itself is an oscillation with small magnitude compared to the idealized universe, so applying the settings in \cite{landau2013fluid}, the SCO causes small changes in ambient density $D_{0}$ and ambient pressure $P_{0}$ in the idealized universe with
          \begin{equation}\label{denpre}
          \begin{aligned}
          &P_{tot}(\textbf{x},t) = P_{0} + P(\textbf{x},t),&\\
          &D_{tot}(\textbf{x},t) = D_{0} + D(\textbf{x},t).
          \end{aligned}
          \end{equation}
        Here $P_{tot}$ and $D_{tot}$ represent the pressure and density of the local medium after being affected by the SCO. $P_{0}$ and $D_{0}$ represent the pressure and density of the medium in equilibrium in the idealized universe without a perturbation. $P$ and $D$ satisfy $P \ll P_{0}$ and $D \ll D_{0}$.
        
       For an adiabatic process, the phase velocity of the plane wave and the pressure and density of the medium have the following relation:
          \begin{equation}\label{adiab}
          c_{s} = \sqrt{ \left(\frac{\partial P_{tot}}{\partial D_{tot}} \right)_{S}}.
          \end{equation}
        The "$S$" refers to the adiabatic process. It follows from Eq.(\ref{adiab}) that the speed of propagation is determined by the pressure and density of the medium. In addition the changes in density and pressure are related as follows:
          \begin{equation}\label{denprerela}
          P = D \left(\frac{\partial P_{tot}}{\partial D_{tot}} \right)_{S}.
          \end{equation}
        Next we need to find a function of $P_{tot}$ with respect to $D_{tot}$, or $D_{tot}$ with respect to $P_{tot}$. For this purpose we first assume that the medium is an ideal gas. So the relation between pressure and density in an ideal gas can be used: 
          \begin{equation}\label{gamma}
          \begin{aligned}
          &P_{tot} = P_{0}\left(\frac{D_{tot}}{D_{0}}\right)^{\gamma_{S}},\\
          &D_{tot} = D_{0}\left(\frac{P_{tot}}{P_{0}}\right)^{\frac{1}{\gamma_{S}}}.
          \end{aligned}
          \end{equation}
        Here $\gamma_{S}$ is the adiabatic exponent with $1 < \gamma_{S} < 2$. In the case when $P_{0} \gg P(\textbf{x},t)$ and $D_{0} \gg D(\textbf{x},t)$, the first-order expansion of Eq.(\ref{gamma}) can be written as:
          \begin{equation}\label{gammaexp}
          \begin{aligned}
          &P_{tot} = P_{0}\left(1 + \frac{D}{D_{0}}\right)^{\gamma_{S}} = P_{0}\left[1 + \gamma_{S}\frac{D}{D_{0}} + \mathcal{O}\left(\left|\frac{D}{D_{0}}\right|^{2}\right) \right] ,&\\
          &D_{tot} = D_{0}\left(1 + \frac{P}{P_{0}}\right)^{\frac{1}{\gamma_{S}}}= D_{0}\left[1 + \frac{1}{\gamma_{S}}\frac{P}{P_{0}} + \mathcal{O}\left(\left|\frac{P}{P_{0}}\right|^{2}\right)  \right].
          \end{aligned}
          \end{equation}
        This means that $P_{0} + (\gamma_{S} P_{0} D/D_{0}) \approx P_{0} + P$ and $D_{0} + (D_{0} P/\gamma_{S} P_{0}) \approx D_{0} + D$. Thus we can obtain the relation: $\gamma_{S} P_{0}/D_{0} \approx P/D$. Now applying Eq.(\ref{adiab}) and Eq.(\ref{denprerela}) we can see that $\sqrt{\gamma_{S} P_{0}/D_{0}}  \approx c_{s,0}$. The result here is $c_{s,0}$, which is the constant part of the propagation speed, because this result is only related to the constants $\gamma_{S}$, $P_{0}$ and $D_{0}$.
        
        Now we apply Eq.(\ref{gamma}) to Eq.(\ref{adiab}) to get the expression for the speed of propagation:
          \begin{equation}\label{propspd}
          \begin{aligned}
          &c_{s} = \sqrt{\frac{\gamma_{S} P_{0}}{D_{0}}  \left(\frac{D_{0}+D}{D_{0}}\right)^{\gamma_{S} - 1} } = c_{s,0}\left(1 + \frac{D}{D_{0}}\right)^{\frac{\gamma_{S} - 1}{2}},&\\
          &c_{s} = \sqrt{\frac{\gamma_{S} P_{0}}{D_{0}}  \left(\frac{P_{0} + P}{P_{0}}\right)^{\frac{\gamma_{S} -1}{\gamma_{S}}} } = c_{s,0}\left(1 + \frac{P}{P_{0}}\right)^{\frac{\gamma_{S} -1}{2\gamma_{S}}}.
          \end{aligned}
          \end{equation}
        The expression with pressure in Eq.(\ref{propspd}) applies a deduction of Eq.(\ref{gamma}): 
          \begin{equation}\label{dedu}
          \begin{aligned}
          \frac{D_{tot}}{D_{0}} = \left(\frac{P_{tot}}{P_{0}}\right)^{\frac{1}{\gamma_{S}}} \rightarrow \left(\frac{D_{tot}}{D_{0}}\right)^{\gamma_{S} - 1} = \left(\frac{P_{tot}}{P_{0}}\right)^{\frac{\gamma_{S} - 1}{\gamma_{S}}}.
          \end{aligned}
          \end{equation}
        Using Eq.(\ref{propspd}) we can also define the refractive index field $n(\textbf{x},t)$ caused by the oscillation:
          \begin{equation}\label{nfield}
          \begin{aligned}
          n(\textbf{x},t) = \frac{c_{s,0}}{c_{s}(\textbf{x},t)} = \left(1 + \frac{D}{D_{0}}\right)^{\frac{1 - \gamma_{S}}{2}},\\
          n(\textbf{x},t) = \frac{c_{s,0}}{c_{s}(\textbf{x},t)} = \left(1 + \frac{P}{P_{0}}\right)^{\frac{1 - \gamma_{S}}{2\gamma_{S}}}.
          \end{aligned}
          \end{equation}
        Here the representation of the refractive index field can be obtained by using either the pressure or density expression of the ideal gas. These two representations of the refractive index field in Eq.(\ref{nfield}) are equivalent. Because the oscillation of an ideal gas causes both the density and pressure of the medium to change, they are two expressions of an oscillation.

        Returning to the equation of motion of SCOs and applying Eq.(\ref{propspd}) or Eq.(\ref{nfield}) to Eq.(\ref{accetime}), as long as we know the initial velocity vector or $\boldsymbol{\beta}_{new}(0)$  of SCO 2, we can express the acceleration of SCO 2 by the oscillation field of SCO 1. Because the effect of SCO 1 on the propagation speed field or refractive index field can now be represented entirely by the oscillation field itself.

    \section{Refractive index field for the inverse-square-type acceleration of two SCOs}    
    \label{sect3}
    
        We start to consider the case where the 2 SCOs are attracted to each other. First of all a simplification of Eq.(\ref{accetime}) is needed because here we only consider the inverse-square-type behavior like the Newtonian gravitation-type acceleration, which holds at a small velocity of the SCOs. So we consider the case where the two SCOs are initially stationary and that $\boldsymbol{\beta}_{new} \ll 1$, i.e., the problem remains in the small velocity regime. We use SCO 1 as a reference system and observe the motion of SCO 2. It can be assumed that the effect of SCO 2 on the local propagation speed field is so small that the acceleration possessed by SCO 1 is also small and can be ignored. This means that SCO 2 is a probe. We can use the spherical coordinate system. Thus the acceleration in Eq.(\ref{accetime}) simplifies to 
          \begin{equation}\label{accetimesim}
          a(r) = -c_{s}(r)\frac{\mathrm{d}c_{s}(r)}{\mathrm{d}r} = -\frac{1}{2} \frac{\mathrm{d}c^{2}_{s}(r)}{\mathrm{d}r}  = \frac{c^{2}_{s,0}}{n^{3}_{ref}(r)} \frac{\mathrm{d}n_{ref}(r)}{\mathrm{d}r}  .
          \end{equation}
        Here the radial component of $\textbf{a}$ is used because the motion of both SCOs follows a straight line connecting the centres of both and acceleration is independent of angle components. The time component $t$ of the acceleration $\textbf{a}$ is not involved in the differential operation, so only $r$, the distance between the centres of the two SCOs, needs to be considered. For convenience, the derivation will be made by using the propagation speed field $c_{s}(r)$ rather than the refractive index field $n_{ref}(r)$, but it is always possible to convert the result to be expressed in terms of the refractive index field by $c_{s}(r) = c_{s,0}/n_{ref}(r)$.
        
        The question we are pondering is: Can SCO 2 have a attractive inverse-square-type acceleration like the Newtonian gravitation-type, $-GM/r^{2}$, where $G > 0$ is a constant and $M > 0$ is a variable associated with SCO 1? If the answer is yes, then the propagation speed $c_{s}(r)$ should satisfy the relation:
          \begin{equation}\label{algebra}
          -\frac{GM}{r^{2}} = -c_{s}(r)\frac{\mathrm{d}c_{s}(r)}{\mathrm{d}r}.
          \end{equation}
        We can therefore derive the expression for $c_{s}(r)$:
          \begin{equation}\label{condi}
          \begin{aligned}
          &\frac{GM}{r^{2}} = c_{s}(r)\frac{\mathrm{d}c_{s}(r)}{\mathrm{d}r} \rightarrow \frac{GM}{r^{2}} \mathrm{d}r = c_{s} \mathrm{d} c_{s} \rightarrow &\\ 
          &c_{s}(r) = \left(h_{1} - \frac{2GM}{r}    \right)^{\frac{1}{2}} 
          \rightarrow n_{ref}(r) = c_{s,0}\left(h_{1} - \frac{2GM}{r}    \right)^{-\frac{1}{2}}.
          \end{aligned}
          \end{equation}
        Here $h_{1} > 0$ is a constant. For convenience we can set $h_{1} = c^{2}_{s,0}$. Thus, if SCO 1 causes the local propagation speed to become $c_{s}(r) =  \sqrt{h_{1} - 2GM/r}  = c_{s,0}\sqrt{1 - 2GM/c^{2}_{s,0}r}$, then SCO 2 will have a inverse-square-type acceleration. This result is also the reason why we introduced the refractive index of an inhomogeneous medium in fluid dynamics, i.e., Section \ref{sec2.2}. We will try to find a propagation speed field that is equal to or approximates the result in Eq.(\ref{condi}) in the next section. 
        
\begin{figure}[H]
	      \centering
		  \includegraphics[width=0.40\textwidth]{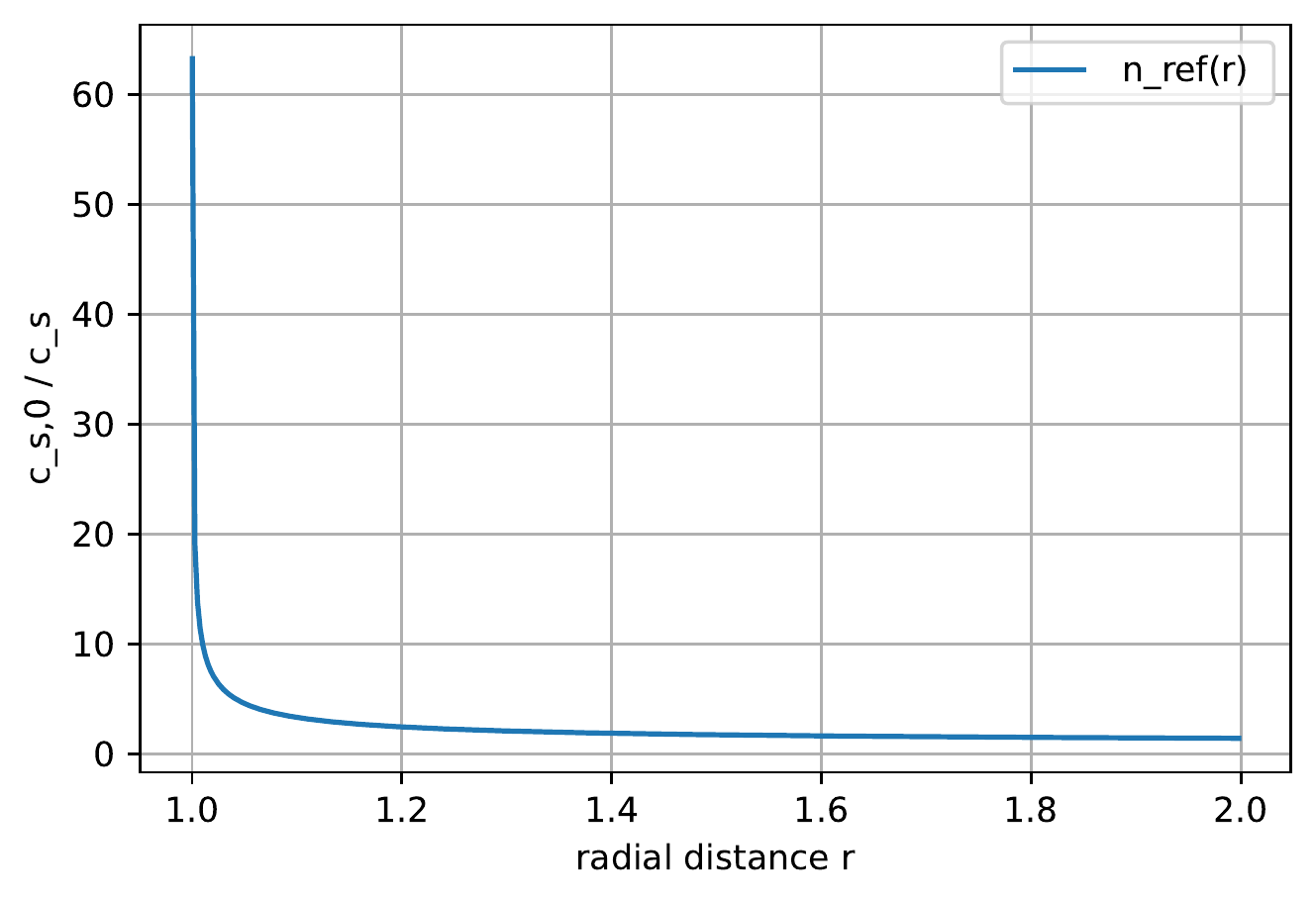}
		  \caption{This figure shows the relation between refractive index and radial distance to the centre of SCO 1: $n_{ref}(r) = c_{s,0} \sqrt{1/(h_{1} - 2GM/r)}$. With $c_{s,0} = 1$, $h_{1} =  1$, $GM = 1/2$.}
		  \label{nref}
          \end{figure}

        The refractive index obtained in Eq.(\ref{condi}) is illustrated in Fig.\ref{nref}. According to Eq.(\ref{condi}), under the condition that the propagation speed is a real value, the value of $r$ should be taken as $r > 2GM/c^{2}_{s,0}$. That is, only when the distance between SCO 1 and SCO 2 is greater than $2GM/c^{2}_{s,0}$, can Eq.(\ref{condi}) be used to describe the acceleration of SCO 2. The refractive index of the medium is high in the region close to the centre of the SCO, which means that the propagation speed is low. The refractive index in the region far from the centre of the SCO tends to be closer to 1.

    \section{Testing SCO models}  
    \label{mathe}
\begin{figure}[H]
	      \centering
		  \includegraphics[width=0.40\textwidth]{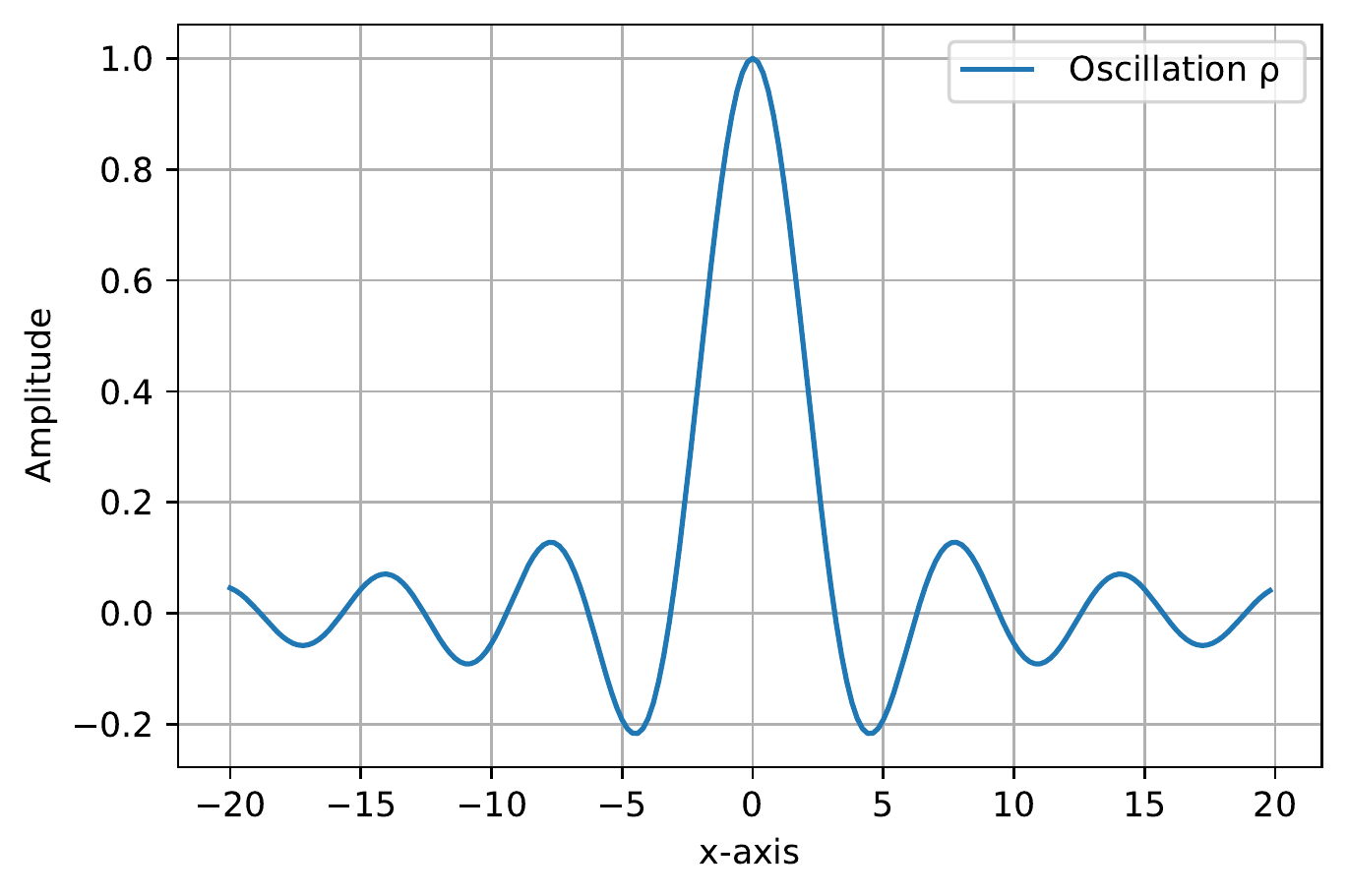}
		  \caption{This figure shows the oscillation (Eq.\ref{spher}) of a single spherical standing wave centred at the origin of the coordinates at time $t = 0$, with $A_{0} = 1$, $k = 1$, $c_{s,0} = 1$. Here any axis through the centre of the SCO can be taken as the x-axis, and due to spherical symmetry this oscillation is of this shape in any direction. }
		  \label{singlefig}
          \end{figure}

        Now we consider the construction of the SCO model. According to Section \ref{SCOModel}, the SCO should be a solution to the 3-dimensional spatial wave equation, Eq.(\ref{waveeq}), so the simplest model of a stationary SCO would be a spherical standing wave. Without loss of generality, using the spherical coordinate system with the centre of SCO 1 as the origin, $\rho(r,t)$ as the oscillation field of SCO 1, we have
          \begin{equation}\label{spher}
          \rho(r,t) = \frac{A_{0}\sin{(kr)}\cos{(kc_{s,0}t)}}{r}.
          \end{equation}
        Here, $A_{0}$ is the factor of amplitude and is constant. $k$ is the wave number. In this case, SCO 1 is an oscillating field centred at the origin, with a period of $2\pi/kc_{s,0}$ and a maximum amplitude of $|A_{0}|$. Fig.\ref{singlefig} illustrates the shape of a single spherical standing wave.
        
        However, it is straightforward to construct a more general form. One SCO can be modeled as a superposition of different spherical standing waves,
          \begin{equation}\label{genespher}
          \begin{aligned}
          \rho(r,t) &= \sum^{k = k_{max}}_{k = k_{min}}  \frac{A(k)\sin{(kr)}\cos{(kc_{s,0}t)}}{r} &\\ 
          &= \sum^{n = N}_{n = 1}  \frac{A(n)\sin{(n\cdot k'r)}\cos{(n\cdot k'c_{s,0}t)}}{r}.
          \end{aligned}
          \end{equation}
        In the model described by Eq.(\ref{genespher}), we assume that the frequencies (and therefore wave numbers) of plane waves that make up the SCO in this idealized universe are quantized by $k = n \cdot k'$. Furthermore, we assume that $k_{max} = N \cdot k'$ where $k'$ is a unit wave number that can change, $N$ and $n$ are positive integers. In our setting, the summation starts at $n = 1$, which means that $k_{min} = k'$. By using the unit wave number $k'$, we can replace $k_{min}$ and $k_{max}$ with positive integers. In the absence of further settings, $A(k)$, $k_{min}$ and $k_{max}$ fully characterize a particular SCO. Finally this also means that $A(k)$, $k'$ and $N$ can fully characterize a particular SCO. $A(k)$ is now a function of the wave number $k$. These settings will facilitate subsequent calculations. We will use the model of Eq.(\ref{genespher}) directly for the rest of the study
    \subsection{Superposition of multiple-spherical standing waves} 
    \label{singlestanwav}
        For the local propagation speed field under the influence of SCO 1: a superposition of multiple-spherical standing waves like Eq.(\ref{genespher}), we first consider the case where $\alpha_{D} \rho(r,t)$ represents the density fluctuation $D(r,t)$. 
    \subsubsection{General form of acceleration of the superposition of multiple-spherical standing waves}   
    \label{aosssw}
        Substituting the oscillation field $\rho(r,t)$ from Eq.(\ref{genespher}) into Eq.(\ref{propspd}), we obtain:
          \begin{equation}\label{densi}
          \begin{aligned}
          c_{s}(r,t) & = c_{s,0}\left[ 1 + \frac{D(r,t)}{D_{0}} \right]^{\frac{\gamma_{S} - 1}{2}} & \\
          & = c_{s,0}\left[ 1 + \sum^{n = N}_{n = 1}  \frac{\alpha_{D}A(n)\sin{(nk'r)}\cos{(nk'c_{s}t)}}{D_{0}r} \right]^{\frac{\gamma_{S} - 1}{2}} 
          \end{aligned}
          \end{equation}
        The plane wave propagation speed in the region around SCO 1 oscillates with time and space.

        In the case $\boldsymbol{\beta}_{new} \ll 1$, SCO 2 will be accelerated under the influence of SCO 1. Notice that the propagation speed is now time dependent. Using Eq.(\ref{accetimesim}) and Eq.(\ref{densi}), after the expansion we have:
          \begin{equation}\label{singleacce}
          \begin{aligned}
          a(r,t) &= \frac{(1 - \gamma_{S})c^{2}_{s,0} }{2}\left[ 1 + \frac{D(r,t)}{D_{0}} \right]^{\gamma_{S} - 2} \left[ \frac{D(r,t)}{D_{0}}  \right]' &\\
          &= \frac{(1 - \gamma_{S})c^{2}_{s,0} }{2} \Bigl\{ 1 + (\gamma_{S} - 2)\frac{D(r,t)}{D_{0}} + &\\ & \quad \mathcal{O}\left[\left|\frac{D(r,t)}{D_{0}} \right|^{2}\right]\Bigr\}\left[ \frac{D(r,t)}{D_{0}}  \right]'&\\
          &= \frac{(1 - \gamma_{S})c^{2}_{s,0}}{2} \Bigl[1 + & \\ 
          & \quad (\gamma_{S} -2) \sum^{n = N}_{n = 1}  \frac{\alpha_{D}A(n)\sin{(nk'r)}\cos{(nk'c_{s,0}t)}}{D_{0}r} &\\
          & \quad +\mathcal{O}\left(\left|\frac{D(r,t)}{D_{0}}\right|^{2}\right)  \Bigr] \times &\\
          & \quad \sum^{m = N}_{m = 1} \Bigl[ \frac{\alpha_{D}A(m)mk'\cos{(mk'r)\cos{(mk'c_{s,0}t)}}}{D_{0}r} &\\
          & \quad -\frac{\alpha_{D}A(m)\sin{(mk'r)\cos{(mk'c_{s,0}t)}}}{D_{0}r^{2}}    \Bigr].
          \end{aligned}
          \end{equation}
        Note the multiplication of the summation terms in Eq.(\ref{singleacce}) . Here we have used 
          \begin{equation}\label{summations}
          \begin{aligned}
          & f = \sum^{n=N}_{n=1} F(n) ,&\\
          & g = \sum^{n=N}_{n=1} G(n) ,&\\
          & f\cdot g = \left[\sum^{n=N}_{n=1} F(n)\right] \left[\sum^{m=N}_{m=1} G(m)\right],
          \end{aligned}
          \end{equation}
        where $F(n)$ and $G(n)$ are arbitrary functions of $n$, $f$ and $g$ are their summations respectively. $m$ is a positive integer.  
          
        Observing Eq.(\ref{singleacce}), we find that the acceleration of SCO 2 varies in direction and magnitude with time. We want a time-independent inverse-square-type acceleration, so it is needed to eliminate the time-dependence of the acceleration.
        
        To this end, we consider the time average of the acceleration $\langle a(r,t) \rangle_{t}$. This means that we integrate the acceleration over a time period and divide it by the length of this time period:
          \begin{equation}\label{timeaver}
          \langle a(r,t) \rangle_{t} = \frac{ \int^{(i+1)T}_{iT} a(r,t) \mathrm{d}t  }{\int^{(i+1)T}_{iT}  \mathrm{d}t}.
          \end{equation}
        Here $T$ is a complete time period of $a(r,t)$ and $i \geq 0$ is an integer. The time average is time independent.
        
        Multiplying the summation terms in Eq.(\ref{singleacce}) together reveals that only terms with factor $\cos^{2}{(nk'c_{s,0}t)}$ can survive after time averaging, i.e., for the terms with $m = n$. The reason for this lies in the setting of the unit wave number $k'$ in Eq.(\ref{genespher}) so that if $m \neq n$, any term in the Eq.(\ref{singleacce}) with factor $\cos{(nk'c_{s,0}t)}\cos{(mk'c_{s,0}t)}$ is time-varying periodically. Further, $2\pi/k'c_{s,0}$ is always a multiple of the period $T$ of the above factor and can therefore be used for the time-averaging calculation. For $m \neq n$, the time average of the term with factor $\cos{(nc_{s,0}t)}\cos{(mc_{s,0}t)}$ results in zero. So the time averaged acceleration is then:
          \begin{equation}\label{accetimeaver}
          \begin{aligned}
          \langle a(r) \rangle_{t} \approx &\frac{(1 - \gamma_{S})(\gamma_{S} - 2)c^{2}_{s,0}\alpha^{2}_{D}}{4D^{2}_{0}} \sum^{n = N}_{n = 1} \Bigl[ \frac{A^{2}(n)nk'\sin{(2nk'r)}}{2r^{2}} &\\
          & -\frac{A^{2}(n)\sin^{2}{(nk'r)}}{r^{3}}   \Bigr].
          \end{aligned}
          \end{equation} 
        With the use of  $\langle \cos^{2}{(t)} \rangle_{t} = \frac{1}{2}$.
        If we want to find a inverse-square-type acceleration, we need to study the function $A(n)$. Different $A(n)$ can affect the form of the acceleration. 
        
        If the oscillation represents a pressure perturbation then:
          \begin{equation}\label{accetimeaverexppre}
          \begin{aligned}
          \langle a(r) \rangle_{t} \approx& \frac{(\gamma_{S} - 1)c^{2}_{s,0}\alpha^{2}_{P}}{4\gamma^{2}_{S}P^{2}_{0}} \sum^{n = N}_{n = 1} \Bigl[ \frac{A^{2}(n)nk'\sin{(2nk'r)}}{2r^{2}} &\\
          & -\frac{A^{2}(n)\sin^{2}{(nk'r)}}{r^{3}}   \Bigr].
          \end{aligned}
          \end{equation}
        Although the sine term is still present in Eq.(\ref{accetimeaver}) or Eq.(\ref{accetimeaverexppre}), the superposition with function $A(n)$ will cancel out the fluctuation caused by the sine term within a space interval. We will show this below.
    \subsubsection{General form of total energy of the superposition of multiple-spherical standing waves}   
    \label{teosssw}
        Now we start to consider the total energy of the SCO. As a solution to the wave equation Eq.(\ref{waveeq}), the energy density $\epsilon$ of the SCO described by Eq.(\ref{genespher}) can be expressed as:
          \begin{equation}\label{singleenerden}
          \begin{aligned}
          \epsilon(r,t) & = \frac{\mu}{2}\left\{ \left[\frac{\partial \rho(r,t)}{\partial t}\right]^{2} + c^{2}_{s,0} [\nabla \rho(r,t)]^{2}    \right\} & \\
          & = \sum^{n = N}_{n = 1}\epsilon_{n}(r,t) + \sum^{m,n = N}_{n \neq m; m,n = i}\epsilon_{n,m}(r,t).
          \end{aligned}
          \end{equation}
        where 
          \begin{equation}\label{singleenerden2}
          \begin{aligned}
          \epsilon_{n}(r,t) &= \mu c^{2}_{s,0} A^{2}(n) \Bigl[ \frac{ n^{2}k'^{2} \sin^{2}{(nk'r)}\sin^{2}{(nk'c_{s,0}t)}   }{2r^{2}} &\\ 
          &\quad +\frac{ n^{2}k'^{2} \cos^{2}{(nk'r)}\cos^{2}{(nk'c_{s,0}t)}   }{2r^{2}} \\
          &\quad + \frac{ \sin^{2}{(nk'r)}\cos^{2}{(nk'c_{s,0}t)}   }{2r^{4}} & \\
          &\quad -\frac{ nk'  \sin{(2nk'r)}\cos^{2}{(nk'c_{s,0}t)}   }{2r^{3}} \Bigr]
          \end{aligned}
          \end{equation}   
        Here $\mu$ is a constant and has units related to the unit of $\rho$. $\mu$ is used to keep the result of $\epsilon$ in the unit of energy density and $\mu > 0$. $n$ and $m$ are positive integers. $\sum^{m,n = N}_{n \neq m; m,n = i}\epsilon_{n,m}(r,t)$ contains all terms that do not have the factor $\cos^{2}{(nk'c_{s,0}t)}$ or $\sin^{2}{(nk'c_{s,0}t)}$, which also means that these terms will disappear if we take the time average of the total energy later, since $\langle \sin{(nk'c_{s,0}t)}\cos{(mk'c_{s,0}t)} \rangle_{t} = \langle \sin{(nk'c_{s,0}t)}\sin{(mk'c_{s,0}t)}\rangle_{t} = \langle \cos{(nk'c_{s,0}t)}\cos{(mk'c_{s,0}t)} \rangle_{t} = 0$ and the time factor is not involved in the volume integral for the total energy. The full expression of $\epsilon_{n,m}(r,t)$ is in Appendix \ref{app1}.
 
        The energy density Eq.(\ref{singleenerden}) is time- and space-dependent.
        The total energy $E$ of the SCO is obtained through volume integration of the energy density:
          \begin{equation}\label{singleener}
          \begin{aligned}
          E = \int_{V} \epsilon(r,t) \mathrm{d}V.
          \end{aligned}
          \end{equation}
        Since the total energy of an unbounded spherical standing wave is infinite, we need to limit the boundaries. The point that a SCO has boundaries was also presented by Stadler et al. in \cite{stadtler2021dynamics}. Furthermore, for the single spherical standing wave with wavelength $\lambda$, only integrals in the range of multiples of half wavelength can make the total energy time-independent. For consistency of the derivation, we set the boundary of the SCO 1 to be $w$ times the half wavelength of the spherical standing wave in advance. Here $w$ is a positive integer, $\lambda/2 = \pi/k$ is the half wavelength. Now for the SCO model of superposition of multiple spherical standing waves described in Eq.(\ref{genespher}), we choose to integrate to $w\pi/k'$, as this is the multiple of the half spatial periods (i.e., the half wavelengths) of all spherical standing wave components in the superposition. The total energy will be:
          \begin{equation}\label{singleenerlim}
          \begin{aligned}
          E_{SCO}(t)  &= 4\pi \int^{\frac{w\pi}{k'}}_{0} \left[\sum^{n = N}_{n = 1}\epsilon_{n}(r,t) + \sum_{n \neq m}\epsilon_{n,m}(r,t)\right] r^{2} \mathrm{d}r &\\ 
          & = \pi^{2}\mu w \sum^{n = N}_{n = 1}  n^{2}k'A^{2}(n) c^{2}_{s,0} + &\\
          & \quad 4\pi \int^{\frac{w\pi}{k'}}_{0} \left[ \sum_{n \neq m}\epsilon_{n,m}(r,t)\right] r^{2} \mathrm{d}r.
          \end{aligned}
          \end{equation}
        Observing Eq.(\ref{singleenerlim}), we find that the total energy is still time-dependent. In order to obtain an expression of the total energy for the SCO that does not vary with time, we need to take the time average of the total energy, which means that the time-dependent terms are cancelled out. So we get:
          \begin{equation}\label{singlewavesingleenerlim}
          \begin{aligned}
          \langle E_{SCO} \rangle_{t}  = \pi^{2}\mu w  \sum^{n = N}_{n = 1}  n^{2}k'A^{2}(n) c^{2}_{s,0}.
          \end{aligned}
          \end{equation}
        Therefore the total energy of the SCO is related to the summation of the $n^{2}A^{2}(n)$ terms.
        If we apply the mass-energy equivalence relation, $E = Mc^{2} $, to the total energy obtained in Eq.(\ref{singlewavesingleenerlim}), then a definition of the mass of the SCO in this idealized universe can be given:
          \begin{equation}\label{singlemass}
          \begin{aligned}
          M_{SCO} = \pi^{2}\mu w \sum^{n = N}_{n = 1}  n^{2}k'A^{2}(n).
          \end{aligned}
          \end{equation}
        In order to find the inverse-square-type acceleration, we need to consider the function $A(n)$. In the following we will show that there is a form of $A(n)$ that allows SCO 2 to have a inverse-square-type acceleration.
        
    \subsubsection{Search for inverse-square-type acceleration}
    \label{sec4.2.3}
        Returning to Eq.(\ref{accetimeaver}) or Eq.(\ref{accetimeaverexppre}), in order to obtain a inverse-square-type acceleration, the superposition term must satisfy:
          \begin{equation}\label{fouri1}
          \begin{aligned}
          \sum^{n = N}_{n = 1} \left[ \frac{A^{2}(n)nk'\sin{(2nk'r)}}{2r^{2}} - \frac{A^{2}(n)\sin^{2}{(nk'r)}}{r^{3}}   \right] = \frac{b}{r^{2}},
          \end{aligned}
          \end{equation}
        where $b$ is a parameter independent of $n$ and the distance $r$ between SCO 1 and SCO 2. The new form of Eq.(\ref{fouri1}) can be obtained after using the transformation $\sin^{2}{(nk'r)} = 1-\cos{(2nk'r)}/2$ and $y = 2k'r$:
          \begin{equation}\label{fouri2}
          \begin{aligned}
          \sum^{n = N}_{n = 1} \left[  \frac{A^{2}(n)ny\sin{(ny)}}{2}  - A^{2}(n) + A^{2}(n)\cos{(ny)}  \right] = \frac{by}{k'} = b'y.
          \end{aligned}
          \end{equation}
        Here we let $b' = b/k'$.
        Observing Eq.(\ref{fouri2}), if we let 
          \begin{equation}\label{fouri3}
          \begin{aligned}
          g(y) = \sum^{n = N}_{n = 1} \left[ A^{2}(n)\cos{(ny)} - A^{2}(n) \right],
          \end{aligned}
          \end{equation}
        then 
          \begin{equation}\label{fouri4}
          \begin{aligned}
          -\frac{y}{2} \left[\frac{\mathrm{d}g(y)}{\mathrm{d}y}\right] = \sum^{n = N}_{n = 1} A^{2}(n)\cdot \frac{ny}{2}\cdot \sin{(ny)},
          \end{aligned}
          \end{equation}
        so Eq.(\ref{fouri2}) can be rewritten as:
          \begin{equation}\label{fouri5}
          \begin{aligned}
          -\frac{y}{2} \left[\frac{\mathrm{d}g(y)}{\mathrm{d}y}\right] + g(y) = b'y.
          \end{aligned}
          \end{equation}
        Solving this ordinary differential Eq.(\ref{fouri5}), we get:
          \begin{equation}\label{fouri6}
          \begin{aligned}
          g(y) = 2b'y + dy^{2},
          \end{aligned}
          \end{equation}
        where $d$ is another parameter independent of $n$ and the distance $r$. Looking at Eq.(\ref{fouri3}) and Eq.(\ref{fouri6}), one finds that in one length period, i.e., $y \in (0, 2\pi)$ or $r \in (0, \pi/k')$, $g(y)$ in Eq.(\ref{fouri6}) can be expanded by a Fourier series for $g(y)$ in Eq.(\ref{fouri3}) if $N\rightarrow \infty$. The general expression of the Fourier series of $g(y)$ is as follows:
          \begin{equation}\label{fouri6ssr}
          \begin{aligned}
          g(y) = \frac{a_{0}}{2} + \sum^{n = N}_{n = 1} [ a_{n}\cos{(ny)} + b_{n}\sin{(ny)} ],
          \end{aligned}
          \end{equation}
        with 
          \begin{equation}\label{fouri6sssr}
          \begin{aligned}
          & a_{0} = \frac{1}{\pi} \int^{2\pi}_{0} g(y) \mathrm{d}y,&\\
          & a_{n} = \frac{1}{\pi} \int^{2\pi}_{0} g(y)\cos{(ny)} \mathrm{d}y,&\\
          & b_{n} = \frac{1}{\pi} \int^{2\pi}_{0} g(y)\sin{(ny)} \mathrm{d}y.
          \end{aligned}
          \end{equation}
        Compare with Eq.(\ref{fouri3}) and Eq.(\ref{fouri6ssr}), the condition that $g(y)$ does not contain sine terms can be used to derive the relation between $d$ and $b'$. It leads to
          \begin{equation}\label{fouri8}
          \begin{aligned}
          \frac{1}{\pi}\int^{2\pi}_{0} (2b'y + dy^{2})\sin{(ny)} \mathrm{d}y = \frac{-4b'}{n} - \frac{4d\pi}{n} =  0,
          \end{aligned}
          \end{equation}
        so $d$ can be expressed in terms of $b'$ as:
          \begin{equation}\label{fouri9}
          \begin{aligned}
          d = -\frac{b'}{\pi}.
          \end{aligned}
          \end{equation}
        Up to now, $g(y)$ has been fully expressed by $b'$ as
          \begin{equation}\label{fourissr}
          \begin{aligned}
          g(y) = 2b'y - \frac{b'y^{2}}{\pi}.
          \end{aligned}
          \end{equation}
        This relation holds for $y \in (0,2\pi)$, i.e., $r \in (0, \pi/k')$. Since we only consider the effective range of the inverse-square-type acceleration, the spatial range of integration required for the total energy of SCO 1 also can be chosen as $(0, \pi/k')$, which means that $w=1$. According to Section \ref{sect3}, in order to enable the existence of inverse-square-type acceleration, there exists a lower bound on $r$ as the distance between SCO 1 and SCO 2, i.e., $r > 2GM/c^{2}_{s,0}$, where $G$ is a constant and $M$ is a variable related to SCO 1. Here we will use the mass of SCO 1, $M_{SCO}$, to represent $M$. A discussion of the lower bound will proceed in Section \ref{discusss}, now we focus on the upper bound of $r$, i.e., $\pi/k'$.
    
        We can now obtain a representation of $A(n)$ by the Fourier series:
          \begin{equation}\label{fouri10}
          \begin{aligned}
          A^{2}(n) = \frac{1}{\pi}\int^{2\pi}_{0} \left(2b'y - \frac{b'}{\pi}y^{2}\right)\cos{(ny)} = -\frac{4b'}{\pi n^{2}}.
          \end{aligned}
          \end{equation}
        Since the energy expression of the general form in Eq.(\ref{singlewavesingleenerlim}) requires the use of $A^{2}(n)$, the condition $A^{2}(n) \geq 0$ should always be satisfied, so there must be $b' \leq 0$ and therefore $b \leq 0$. We therefore obtain a representation of the function $A(n)$ at which the local propagation speed field induced by SCO 1 will cause SCO 2 to acquire a inverse-square-type acceleration in the range $r \in (2GM/c^{2}_{s,0}, \pi/k')$. The $A(n)$ is given as:
          \begin{equation}\label{fouri11}
          \begin{aligned}
          A(n) = \frac{2}{n}\sqrt{\frac{-b}{\pi k'}}.
          \end{aligned}
          \end{equation}
        
\begin{figure}[H]
	      \centering
		  \includegraphics[width=0.40\textwidth]{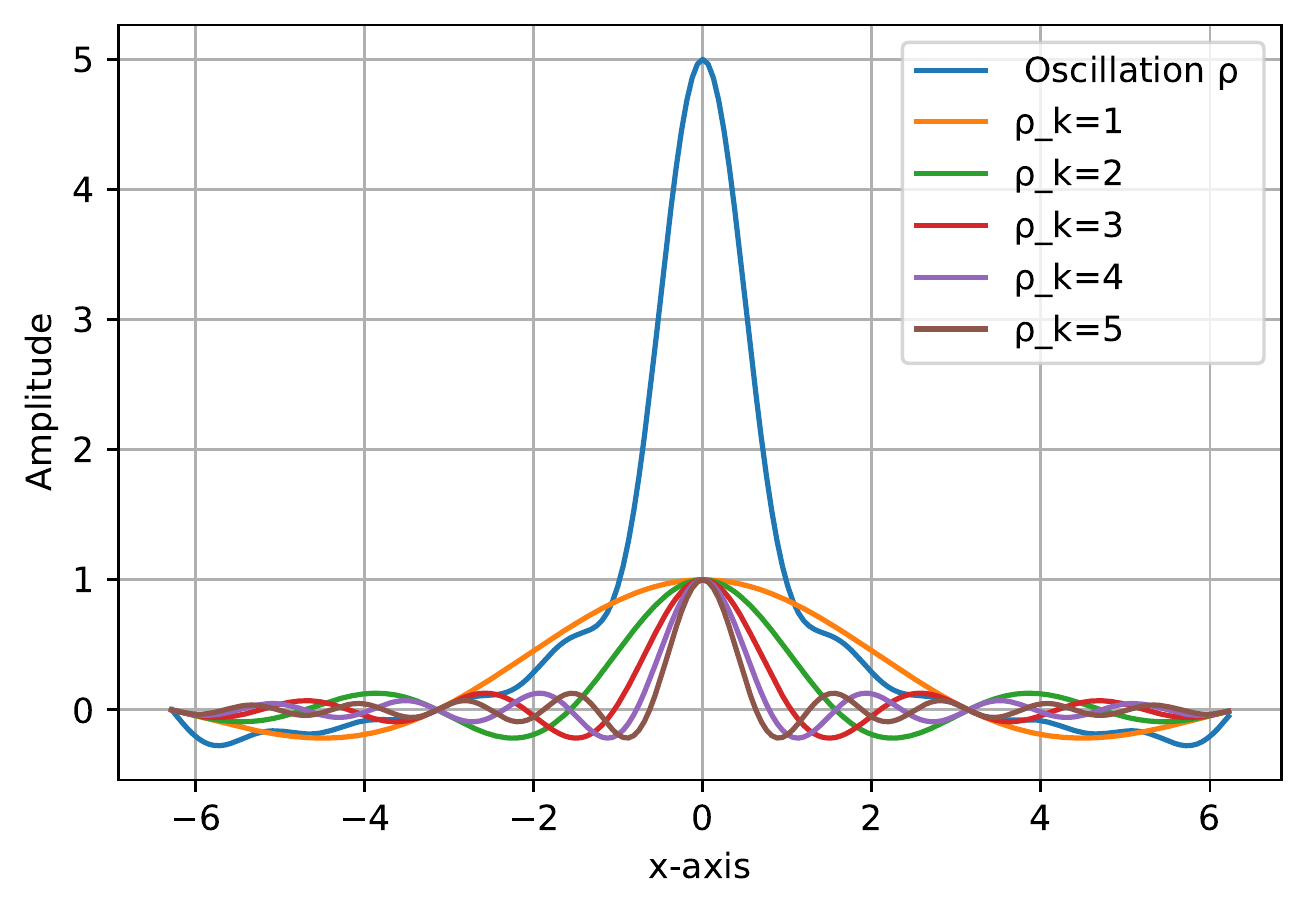}
		  \caption{This figure shows the oscillation $\rho(r,t) = \sum^{n = 5}_{n = 1}  A(n)\sin{(n\cdot k'r)}\cos{(n\cdot k'c_{s,0}t)}/r$ centred at the origin of the coordinates at time $t = 0$ in the range $r\in(0, 2\pi)$. With $A(n) = 1/nk'$, $k' = 1$, $c_{s,0} = 1$. "Oscillation $\rho$" represents the oscillation of 5 spherical standing waves superimposed. "$\rho_{k = i}$" is the oscillation of 5 individual spherical standing waves centred at the origin with $A(n)$ and different wave number $k = nk' =i$, respectively.}
		  \label{5fig}
          \end{figure}

        An SCO with a function $A(n)$ according to Eq.(\ref{fouri11}) is shown in Fig.\ref{5fig}.
        For large N, the SCO  will have a shape as shown in Fig.\ref{40fig}:
          \begin{figure}[H]
	      \centering
		  \includegraphics[width=0.40\textwidth]{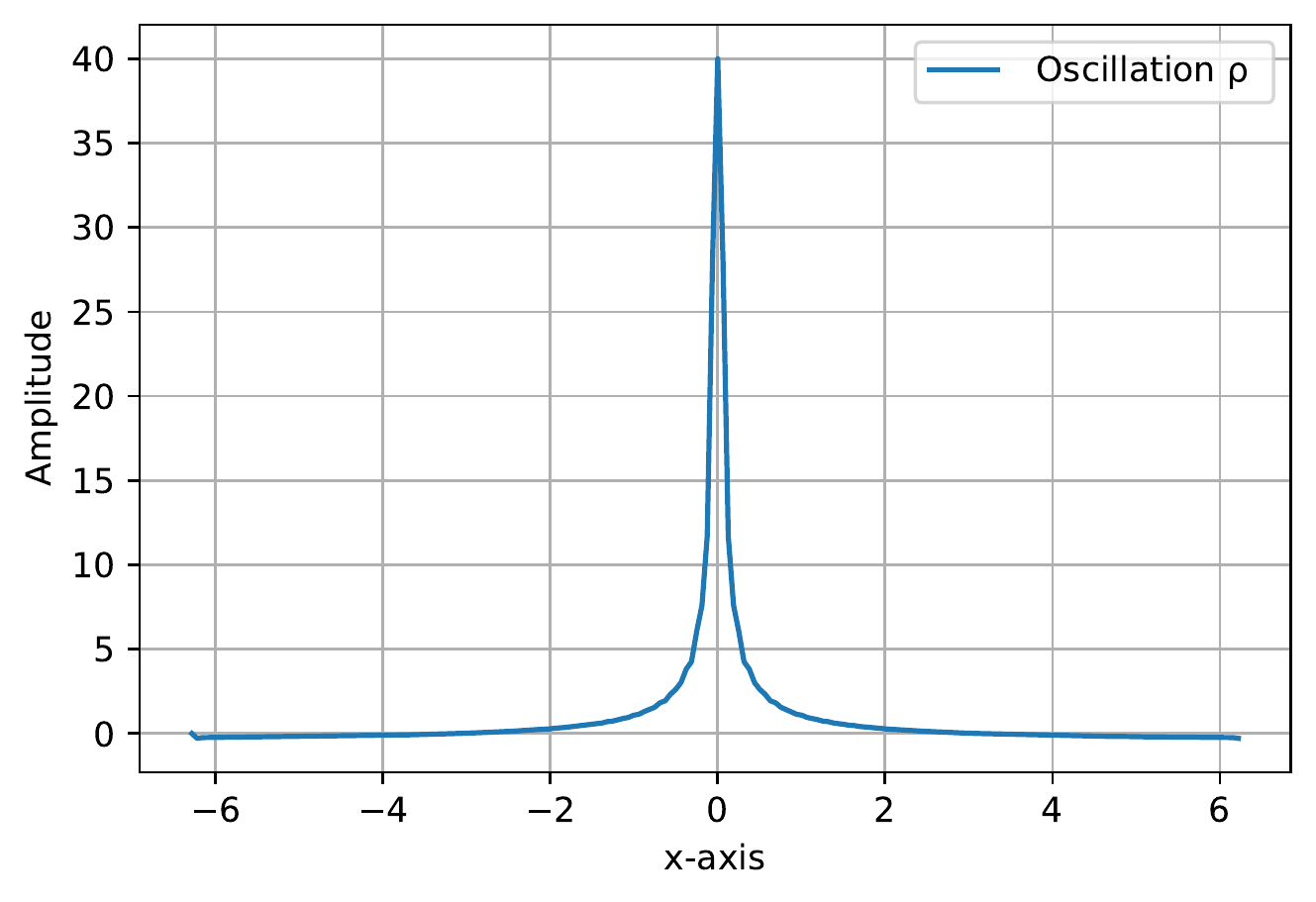}
		  \caption{This figure shows the oscillation $\rho(r,t) = \sum^{n = 40}_{n = 1}  A(n)\sin{(n\cdot k'r)}\cos{(n\cdot k'c_{s,0}t)}/r$ centred at the origin of the coordinates at time $t = 0$ in the range $r\in(0, 2\pi)$. With $A(n) = 1/nk'$, $k' = 1$, $c_{s,0} = 1$. "Oscillation $\rho$" represents the oscillation of the 40 spherical standing waves superimposed. This figure can be used to compare with the single spherical standing wave model in Fig.\ref{singlefig}.}
		  \label{40fig}
          \end{figure}
        It can be seen from Fig.\ref{40fig} that the central peak of the oscillation will become higher and the rest of the secondary peaks will be smoothed out.

        Taking this result Eq.(\ref{fouri11}) back to equation Eq.(\ref{fouri2}) and Eq.(\ref{accetimeaver}), we get a acceleration for the case of the density oscillation:
          \begin{equation}\label{fouri12}
          \begin{aligned}
          \langle a(r) \rangle_{t} &\approx \frac{(1 - \gamma_{S})(\gamma_{S} - 2)c^{2}_{s,0}\alpha^{2}_{D}}{8D^{2}_{0}r^{3}} \sum^{n = N}_{n = 1} \Bigl[ \frac{A^{2}(n)ny \sin{(ny)}}{2} &\\
          &\quad - A^{2}(n) + A^{2}(n)\cos{(ny)}  \Bigr] &\\
          & = \frac{(1 - \gamma_{S})(\gamma_{S} - 2)c^{2}_{s,0}\alpha^{2}_{D}b}{4D^{2}_{0}r^{2}}.
          \end{aligned}
          \end{equation}
        Thus, a inverse-square-type acceleration is obtained if the conditions that the adiabatic exponent $\gamma_{S} \in (1, 2)$ and $b < 0$ hold. According to the description of the adiabatic exponent in Section \ref{sec2.2}, as long as the medium is an ideal gas, it needs to satisfy the condition that $1 < \gamma_{S} < 2$, so when $b < 0$ there is a attractive  acceleration, i.e., SCO 2 always has an acceleration towards SCO 1.
        
        Similarly, in the case of a pressure oscillation we have:
          \begin{equation}\label{fouri13}
          \begin{aligned}
          \langle a(r) \rangle_{t} \approx \frac{(\gamma_{S} - 1)c^{2}_{s,0}\alpha^{2}_{P}b}{4\gamma^{2}_{S}P^{2}_{0}r^{2}}.
          \end{aligned}
          \end{equation}
        In this case, a inverse-square-type acceleration also exists as long as $b < 0$. 
        
        The upper bound of the distance $r$, i.e., $\pi/k'$, is sufficiently long because $k' = k_{min}$, which is the minimum wave number and the maximum wavelength that a spherical standing wave superimposed into the SCO 1 can have.  Furthermore, it can be assumed that the idealized universe in which the SCO exists is spatially finite, such that the wavelegth $\lambda' = 2\pi/k'$ and the spatial scale of the universe is of the same order. In this way, inverse-square-type acceleration would exist on a large spatial scale.
        
        From both representations of the acceleration, we find that $b$ is the only variable parameter apart from the distance $r$. We will therefore use the parameter $b$ to construct the expression of mass in the idealized universe later.
        
        The results in Eq.(\ref{fouri12}) and Eq.(\ref{fouri13}) can be derived directly from Eq.(\ref{fouri1}) and can be verified by substituting the expression $A^{2}(n)$ from Eq.(\ref{fouri10}) into them, this requires the use of the following mathematical relations:
          \begin{equation}\label{math1}
          \begin{aligned}
          \sum^{\infty}_{n = 1} \frac{\sin{(nr)}}{n} &= \frac{\pi - r}{2}, r\in [0,2\pi], \\& \\
          \sum^{\infty}_{n = 1} \frac{\cos{(nr)}}{n^{2}} &= \frac{r^{2}}{4} - \frac{\pi r}{2} + \frac{\pi^{2}}{6}, r\in [0,2\pi].
          \end{aligned}
          \end{equation}

      \subsubsection{Possible non inverse-square-type behavior}
      \label{sectmond}  
        As stated in the introduction, to explain the astronomical phenomenon of mismatched Newtonian gravitation, Milgrom in 1983 proposed a modification of the gravitation theory \cite{Milgrom1983AMO}. In this theory, objects with a low acceleration (In other words, it is affected by a weak gravitation) would be subjected to a different gravitation than Newtonian gravitation. There exist some expressions for the modified Newtonian acceleration of gravitation, one effective form could be (Bekenstein and Milgrom \cite{Bekenstein1984DoesTM})
		
		  \begin{equation}\label{MOND}
          \begin{aligned}
          g = -\frac{G_{N}M}{r^{2}} - \frac{\sqrt{G_{N}Ma_{0}}}{r}.
          \end{aligned}
          \end{equation}
        Here $G_{N}$ is the gravitational constant. $g$ is the gravitational acceleration. $a_{0}$ is a constant, and when $a_{0} \gg G_{N}M/r$ (in other words, when $r$ is very large for constant $G_{N}$ and $M$), the $1/r$ term dominates the acceleration through gravitation, i.e., this could be the MONDian regime. When $a_{0} \ll G_{N}M/r$ (when $r$ is very small for constant $G_{N}$ and $M$), the $1/r^{2}$ term dominates the acceleration through gravitation, i.e., this is the Newtonian regime. Thus the MOND theory expresses a non inverse-squared-type interaction.
        
        We can also try to construct this kind of non inverse-square-type behaviour in the SCO model.
        For this purpose we will modify Eq.(\ref{fouri1}) to
          \begin{equation}\label{fourimond1}
          \begin{aligned}
          \sum^{n = N}_{n = 1} \left[ \frac{A^{2}(n)nk'\sin{(2nk'r)}}{2r^{2}} - \frac{A^{2}(n)\sin^{2}{(nk'r)}}{r^{3}}   \right] = \frac{b_{1}}{r} + \frac{b_{2}}{r^{2}}.
          \end{aligned}
          \end{equation}
        $b_{1}$ and $b_{2}$ are constant values. This means that at very small $r$ ($r \ll 1$ in the unit system) the acceleration of SCO 2 is dominated by the $1/r^{2}$ term and at very large $r$ ($r \gg 1$ in the unit system) the acceleration of SCO 2 is dominated by the $1/r$ term. However a further derivation shows that the $1/r$ term and the $1/r^{2}$ term cannot coexist in the interval $(0,\pi/k')$. The detailed proof is in the Appendix \ref{appA}. 

    \subsubsection{Total energy and mass under the inverse-square-type model}   
    \label{sec4.2.4}
        Substituting the result of $A(n)$ in Eq.(\ref{fouri11}) back into Eq.(\ref{singlewavesingleenerlim}) and using $w=1$, we obtain a representation of the total energy of the SCO under this model:
          \begin{equation}\label{fouri14}
          \begin{aligned}
          \langle E_{SCO} \rangle_{t}  &= -\pi^{2}\mu c^{2}_{s,0} \sum^{n = N}_{n = 1}  \left(n^{2}k'\cdot \frac{4b}{\pi n^{2}k'}\right)&\\
          &= -4\pi\mu b N c^{2}_{s,0}.
          \end{aligned}
          \end{equation}
        The corresponding mass is expressed as:
          \begin{equation}\label{fouri15}
          \begin{aligned}
          M_{SCO} = -4\pi\mu b N.
          \end{aligned}
          \end{equation}
        So the mass or energy of SCO 1 is related to the value of $b$ and number of spherical standing waves $N$ that make up SCOs. As mentioned in Section \ref{sec4.2.3}, $b \leq 0$, which ensures that the total mass $M_{SCO}$ is not negative.
          
        \subsubsection{Further discussion of the parameters}  
        \label{discusss}
        Observing Eq.(\ref{fouri12}) and Eq.(\ref{fouri13}), we find that the acceleration possessed by SCO 2 is related to the value of $b$. $b$ is also the only variable in these representations that can characterize SCO 1, i.e., only the value of $b$ can be used to distinguish between different SCO 1s. The rest of the parameters are global ambient values. Then observing the expression of energy in Eq.(\ref{fouri14}) or mass in Eq.(\ref{fouri15}), we find that energy and mass are related to $b$ as well as $N$. 
        
        In fact the choice of $b$ is completely free in the absence of further settings, and $b$ can be a constant or a variable independent of the distance $r$. As stated at the beginning of Section \ref{mathe}, without further settings $A(k)$, $k'$ and $N$ can all be used to characterize an SCO 1. Here $b$ represents the contribution of $A(k)$ via Eq.(\ref{fouri11}).
        
        The discussion of $k'$ is given in Section \ref{sec4.2.3}.  It mainly affects the effective range of inverse-square-type interactions caused by the SCO. Since $k'$ can be chosen without affecting the acceleration and total energy of the SCO, it is always possible to choose a suitable $k'$ such that the interaction between the SCOs is a long-range interaction.
        
        Now we also need to discuss the choice of $N$.  According to the derivation in Section \ref{sec4.2.3}, in order to satisfy the Fourier series relation, $N$ must tend to infinity. This means that $N$ can no longer be applied to characterize SCO 1, but can be considered as a global parameter unrelated to a particular SCO. According to the derivation in Section \ref{sec4.2.4}, to avoid the energy tending to infinity, we need to absorb $N$. For this we can assume the constant $\mu = \mu_{0}/N$ in the unit system. That is, let $\mu$ be numerically equal to $1/N$ and $\mu_{0}$ be the unit value. In this way whether $N$ tends to infinity or not does not affect the total energy. Then we have:
          \begin{equation}\label{fouri14energy}
          \begin{aligned}
          \langle E_{SCO} \rangle_{t} &= -4\pi\mu_{0} b c^{2}_{s,0}, &\\
          M_{SCO} &= -4\pi\mu_{0} b. &
          \end{aligned}
          \end{equation}
        That is, the value $b$ from the funtion $A(k)$ characterizes the energy and mass of SCO 1. Another option is to make $N$ a very large constant, which also prevents the energy of the SCO from going to infinity and approximately gives the above result in Eq.(\ref{fouri14energy}). This is because the derivation of the Fourier series part in Section \ref{sec4.2.3} would then only hold approximately.
        
        Since both acceleration and mass are known, a comparison with Eq.(\ref{algebra}) and Eq.(\ref{fouri12}) or Eq.(\ref{fouri13}) leads to an expression for $G$. In the range over which the inverse-square-type acceleration is valid, we might be able to compare $G$ with the gravitational constant $G_{N}$. For the case of density oscillations we have:
          \begin{equation}\label{fouri16sss}
          \begin{aligned}
          &-\frac{GM_{SCO}}{r^{2}} = \frac{(1-\gamma_{S})(\gamma_{S}-2)c^{2}_{s,0}\alpha^{2}_{D}b}{4D^{2}_{0}r^{2}} &\\
          &\rightarrow G = \frac{(1-\gamma_{S})(\gamma_{S}-2)c^{2}_{s,0}\alpha^{2}_{D}}{16\pi\mu_{0} D^{2}_{0}}.
          \end{aligned}
          \end{equation}
          
        Similarly, for the case of pressure oscillations,
          \begin{equation}\label{fouri16sssr}
          \begin{aligned}
          G = \frac{(\gamma_{S}-1)c^{2}_{s,0}\alpha^{2}_{P}}{16\pi\mu_{0} \gamma^{2}_{S} P^{2}_{0}},
          \end{aligned}
          \end{equation}
        Since $1 < \gamma_{S} < 2$ and $\mu > 0$, the value of $G$ is greater than zero in both cases of oscillation, this means that the SCO interactions in our model do lead to mutual attraction. Furthermore, we find that $G$ is indeed only related to the constants of the ambient medium, such as ambient density, ambient pressure, adiabatic exponent, etc. $G$ is a universal constant in this idealized universe.
        
        Since $G$ and $M_{SCO}$ are known, we can also obtain the expression of the lower bound on the distance $r$ between SCO 1 and SCO 2 mentioned in Section \ref{sect3}. An inverse-square-type acceleration exists only if the distance $r > 2GM_{SCO}/c^{2}_{s,0}$ with
          \begin{equation}\label{relation01}
          \begin{aligned}
          \frac{2GM_{SCO}}{c^{2}_{s,0}} = \frac{(\gamma_{S}-1)(\gamma_{S}-2)\alpha^{2}_{D}b}{2 D^{2}_{0}}
          \end{aligned}
          \end{equation}
        for the case of density oscillation, and 
          \begin{equation}\label{relation02}
          \begin{aligned}
          \frac{2GM_{SCO}}{c^{2}_{s,0}} = \frac{(1-\gamma_{S})\alpha^{2}_{P}b}{2 \gamma^{2}_{S} P^{2}_{0}}
          \end{aligned}
          \end{equation}
        for the case of pressure oscillation.

        Noting the upper bound on $r$ derived in Section \ref{sec4.2.3}, i.e., $r < \pi/k'$, we can obtain the relation 
          \begin{equation}\label{relation}
          \begin{aligned}
          \frac{2GM_{SCO}}{c^{2}_{s,0}} < \frac{\pi}{k'},
          \end{aligned}
          \end{equation}
        which gives the upper bound on the choice of $k'$. For the case of density oscillation, it is
          \begin{equation}\label{relation2}
          \begin{aligned}
          k' < \frac{2\pi D^{2}_{0}}{(\gamma_{S}-1)(\gamma_{S}-2)\alpha^{2}_{D}b}.
          \end{aligned}
          \end{equation}
        And for the case of pressure oscillation, it is
          \begin{equation}\label{relation3}
          \begin{aligned}
          k' < \frac{2\pi \gamma^{2}_{S} P^{2}_{0}}{(1-\gamma_{S})\alpha^{2}_{P}b}.
          \end{aligned}
          \end{equation}
        Since both $b$ and $1-\gamma_{S}$ are negative, the upper bound on $k'$ remains positive as well. The $k'$ neither impacts the mass of the SCO nor is it a function of $b$. So the value of $k'$ can be taken to be much smaller than this upper bound, which can make the effective range $(2GM_{SCO}/c^{2}_{s,0}, \pi/k')$ of the inverse-square-type interaction arbitrarily wide.

        Furthermore, according to Section \ref{sec2.2}, the pressure expression in Eq.(\ref{fouri16sssr}) and the density expression in Eq.(\ref{fouri16sss}) for G are equivalent. Thus the relation between the constants $\alpha_{D}$ and $\alpha_{P}$ can be obtained:
          \begin{equation}\label{add1}
          \begin{aligned}
          &\frac{(1-\gamma_{S})(\gamma_{S}-2)c^{2}_{s,0}\alpha^{2}_{D}}{16\pi\mu_{0} D^{2}_{0}} = \frac{(\gamma_{S}-1)c^{2}_{s,0}\alpha^{2}_{P}}{16\pi\mu_{0} \gamma^{2}_{S} P^{2}_{0}} & \\
          &\rightarrow \frac{\alpha^{2}_{P}}{\alpha^{2}_{D}} = \frac{(2-\gamma_{S})\gamma^{2}_{S}P^{2}_{0}}{D^{2}_{0}}.
          \end{aligned}
          \end{equation}

    \section{Conclusion}  
    
        We develop a model of a matter particle as a SCO in an idealized universe in which the existence of a classical ideal gas as a medium is assumed. The SCO as a particle is modelled as an oscillation of the medium and one possible mathematical structure of the SCO is described in Section \ref{mathe} and Eq.(\ref{genespher}). Under the above assumptions, using SCO 1 as a reference system, SCO 2 can have a attractive inverse-square-type acceleration towards SCO 1 in the range $r \in (2GM_{SCO}/c^{2}_{s,0}, \pi/k')$ where $\pi/k'$ can be set to be comparable to the size of the idealized universe, and vice versa. The mass expression for the SCO, $M_{SCO}$, and the constant $G$, which can be used for comparison with the gravitational constant $G_{N}$ in this inverse-square-type acceleration, can then be obtained by using the mass-energy equivalence. 
        
        The central idea of this model is the use of time averaging and component superposition. This means that the attraction we get under this model is a macroscopic result. 
        This may imply the possibility that other types of interactions at the microscopic scale can lead to inverse-square-type interactions at the macroscopic scale.
        Furthermore, this contribution discusses the possibility that a macroscopic inverse-square-type interaction originates from the wave nature of matter. 
         
        The constant term from the inverse-square-type interaction can be expressed entirely in terms of the parameters of the ambient medium (i.e., the vacuum or the historical "aether"). By further exploration, it may be possible to obtain more ambient medium representations of constants from other kinds of interactions, such as interactions that are proportional to the power of the distance, or interactions that are exponential to the distance.

        Imagine two extremely large regions, but with different properties of the ambient medium, which may lead to differences in the measurement of these constants in the two regions. This also hints at the possibility that these universal "constants" are not constant, and that the values currently measured are local values.

        This discussion is limited to the classical wave equation and indicates that two SCOs can attract each other in terms of a inverse-square-type effect. A self-consistent treatment will need to take into account a non-classical wave equation that allows $c_s$ to be variable. This leads to the view that SCOs would be solitons (e.g., Rajaraman \cite{rajaraman1982solitons}).

        This model is still in the early stages of research. We have used only the ideal classical gas model as an "aether". The "aether" and Michelson-Morley experiment are discussed in Schmid and Kroupa \cite{schmid_kroupa_2014}. In future research, attempts can be made to extend the medium in the model to superfluids (Landau and Lifshitz\cite{landau2013fluid}).
        
        On the other hand, other models for the mutual attraction of standing waves or wave packets in fluids exist, such as the secondary Bjerknes force (Bjerknes \cite{bjerknes1906fields}). This will also be one of the directions of our further research.

    \section{Data availability}  
        Data generated or analyzed during this study are provided in full within the published article.
        
    \section{Competing interest}  
        The authors declare there are no competing interests.

		\printbibliography

	\appendix
	\section{The full expression of non-squared terms}
	\label{app1}
	    In this section we express in full the term $\epsilon_{n,m}(r,t)$ in Eq.(\ref{singleenerden})
	      \begin{equation}\label{multiwavesingleenerden2s}
          \begin{aligned}
          \epsilon_{n,m}(r,t) & =  \mu c^{2}_{s,0} \Bigl\{ \Bigl[ \frac{A(n)A(m)mnk'^{2}}{r^{2}}  \sin{(nk'r)}\sin{(mk'r)} \times \\ & \quad \sin{(nk'c_{s,0}t)} \sin{(mk'c_{s,0}t)}            \Bigr] + &\\
          &\quad  \Bigl[ \frac{A(n)A(m)mnk'^{2}}{r^{2}}  \cos{(nk'r)}\cos{(mk'r)} \times \\ & \quad \cos{(nk'c_{s,0}t)} \cos{(mk'c_{s,0}t)}            \Bigr] - &\\
          &\quad  \Bigl[ \frac{A(n)A(m)nk'}{r^{3}}  \cos{(nk'r)}\sin{(mk'r)} \times \\ & \quad \cos{(nk'c_{s,0}t)} \cos{(mk'c_{s,0}t)}            \Bigr] - &\\
          &\quad  \Bigl[ \frac{A(n)A(m)mk'}{r^{3}}  \cos{(mk'r)}\sin{(nk'r)} \times \\ & \quad \cos{(nk'c_{s,0}t)} \cos{(mk'c_{s,0}t)}            \Bigr] + &\\
          &\quad  \Bigl[ \frac{A(n)A(m)}{r^{4}}  \sin{(nk'r)}\sin{(mk'r)} \times \\ & \quad \cos{(nk'c_{s,0}t)} \cos{(mk'c_{s,0}t)}            \Bigr] \Bigr\}.
          \end{aligned}
          \end{equation} 
        Here $\mu$  is a constant related to the unit of $\rho$. $\rho$ is the oscillation field. $c_{s,0}$ is the propagation speed of plane waves in the ambient medium. $n$ and $m$ are positive integers and $n \neq m$. If we take the time average of $\epsilon_{n,m}(r,t)$ with a time period $T = 2\pi/k'c_{s,0}$, the result will be zero.

	\section{Search for non inverse-square-type acceleration}
	\label{appA}
	    For the acceleration of the SCO 2 described in Section \ref{sectmond}, we have
	      \begin{equation}\label{fourimond2}
          \begin{aligned}
          \sum^{n = N}_{n = 1} \left[ \frac{A^{2}(n)nk'\sin{(2nk'r)}}{2r^{2}} - \frac{A^{2}(n)\sin^{2}{(nk'r)}}{r^{3}}   \right] = \frac{b_{1}}{r} + \frac{b_{2}}{r^{2}}.
          \end{aligned}
          \end{equation}
        Here $b_{1}$ and $b_{2}$ are parameters that are independent of the distance $r$.
        Let $y = 2k'r$ and perform the same transformation as in Eq.(\ref{fouri2}). Now we have:
          \begin{equation}\label{fourimond3}
          \begin{aligned}
          \sum^{n = N}_{n = 1} \left[  \frac{A^{2}(n)ny\sin{(ny)}}{2}  - A^{2}(n) + A^{2}(n)\cos{(ny)}  \right] =  b'_{1}y^{2} + b'_{2}y.
          \end{aligned}
          \end{equation}
        Here we use $b'_{1} = b_{1}/2k'^{2}$ and $b'_{2} = b_{2}/k'$. Then we let
          \begin{equation}\label{fourimond4}
          \begin{aligned}
          g(y) = \sum^{n = N}_{n = 1} \left[ A^{2}(n)\cos{(ny)} - A^{2}(n) \right].
          \end{aligned}
          \end{equation}
        Substituting this result into Eq.(\ref{fourimond3}), similar to Eq.(\ref{fouri5}) we will obtain an ordinary differential equation:
          \begin{equation}\label{fourimond5}
          \begin{aligned}
          -\frac{y}{2} \left[\frac{\mathrm{d}g(y)}{\mathrm{d}y}\right] + g(y) = b'_{1}y^{2} + b'_{2}y.
          \end{aligned}
          \end{equation}
        Solving this ordinary differential equation, we have:
          \begin{equation}\label{fourimond6}
          \begin{aligned}
          g(y) = -2b'_{1}y^{2}\ln{(y)} + 2b'_{2}y + dy^{2},
          \end{aligned}
          \end{equation}
        here $d$ is an arbitrary constant. Comparing Eq.(\ref{fourimond6}) with Eq.(\ref{fourimond4}), we find that one condition should be satisfied: The $g(y)$ expressed in Eq.(\ref{fourimond6}) should be symmetric along the axis of $y_{0} = \pi$ when the domain of the function $g(y)$ is $y \in (0, 2\pi)$. That is, 
          \begin{equation}\label{fourimond7}
          \begin{aligned}
          g(y_{0} + y') = g(y_{0} - y').
          \end{aligned}
          \end{equation}
        Here $y_{0} = \pi$ and $y = y_{0} + y'$. It holds when $y' \in [0, \pi)$.
        Substituting the $g(y)$ from Eq.(\ref{fourimond6}) into Eq.(\ref{fourimond7}), we have:
          \begin{equation}\label{fourimond8}
          \begin{aligned}
          b'_{1}[  (\pi + y')^{2} \ln{(\pi + y')} - (\pi - y')^{2} \ln{(\pi - y')}  ] = 2b'_{2}y' + 2\pi dy'.
          \end{aligned}
          \end{equation}
        Taking the derivative of $y'$ twice on both sides, we find
          \begin{equation}\label{fourimond9}
          \begin{aligned}
          2b'_{1}[ \ln{(\pi + y')} - \ln{(\pi - y')} ] = 0.
          \end{aligned}
          \end{equation}
        For arbitrary $b'_{1}$ the only real-valued solution is $y' = 0$, but this kind of acceleration in Eq.(\ref{fourimond2}) is expected to hold on $y \in(0, 2\pi)$, i.e., $y' \in [0, \pi)$. So $b'_{1} \equiv 0$ is the only choice. Considering $b'_{1} = b_{1}/2k'^{2}$, we find that the $1/r$ term in the acceleration of the SCO 2 must vanish.

	\end{multicols}
\end{document}